\documentclass[12pt]{article}
\usepackage{geometry,amsmath,amssymb,epsfig}
\newcommand{\emdash}{---}
\newcommand{\mathe}{\mathrm{e}}
\newcommand{\tmop}[1]{\ensuremath{\operatorname{#1}}}

\catcode`\@=11
\@addtoreset{equation}{section}

\catcode`\@=12
\begin{document}
\title{Cluster simulation of relativistic fermions in two space-time
dimensions}

\author{
Ulli Wolff\thanks{
e-mail: uwolff@physik.hu-berlin.de} \\
Institut f\"ur Physik, Humboldt Universit\"at\\ 
Newtonstr. 15 \\ 
12489 Berlin, Germany
}
\date{}

\maketitle

\begin{abstract}
For Majorana-Wilson lattice fermions in two dimensions we derive
a dimer representation. This is equivalent to
Gattringer's loop representation, but is made exact here on
the torus. A subsequent dual  mapping
leads to yet another representation in which a highly efficient 
Swendsen-Wang type cluster
algorithm is constructed.
It includes the possibility of fluctuating boundary conditions.
It also allows for improved estimators
and makes interesting new observables accessible to Monte Carlo.
The algorithm is compatible  with the Gross-Neveu
as well as an additional Z(2) gauge interaction. In this article
numerical demonstrations are reported for critical free fermions.
\end{abstract}
\begin{flushright} HU-EP-07/25 \end{flushright}
\begin{flushright} SFB/CCP-07-38 \end{flushright}
\thispagestyle{empty}
\newpage

\section{Introduction}

Occasionally in talks or papers about dynamical fermions it is mentioned
{\emdash} more or less as a joke {\emdash} that the computer has no data-type
Grassmann and one hence can simulate fermions only via the nonlocal effective
theory after integrating them into the determinant. Of course, this is plagued
by the well-known inefficiencies. In this article, based on Gattringer's loop
representation {\cite{Gattringer:1998cd}}, we show that in two space-time
dimensions one actually can get pretty close to `simulating Grassmann
numbers'{\footnote{Cum grano salis.}}. We here expand on
{\cite{Gattringer:1998cd}} and its recent numerical implementation
{\cite{Gattringer:2007em}} in several ways. First the loop representation is
re-derived starting from Majorana fermions in what we think is a particularly natural
way. The new connection includes definite boundary conditions on the torus and
does not only work in the thermodynamic limit as before. In particular, we can
then also approach the finite volume continuum limit. Furthermore we propose a
cluster algorithm that is (practically) free of critical slowing down and
allows for improved estimators. In this formulation we can simulate
fluctuating boundary conditions which is necessary to allow for fixed
(anti)periodic boundary conditions in the original fermion system. It also
makes ratios of partition functions accessible as observables in Monte Carlo
simulations. They constitute interesting quantities in the continuum limit.

The original Gross-Neveu model of self-coupled fermions in two dimensions
{\cite{Gross:1974jv}} is most naturally written in terms of $N$ species of
Majorana fermions. In the lattice discretization with Wilson fermions the
euclidean action is given by {\cite{Korzec:2006hy}}
\begin{equation}
  S = a^2 \sum_x \left\{  \frac{1}{2} \xi^{\top}_{} \mathcal{C}(\gamma_{\mu}
  \tilde{\partial}_{\mu} + m - \frac{r}{2} a \partial^{\ast} \partial) \xi -
  \frac{g^2}{8} ( \xi^{\top} \mathcal{C} \xi)^2 \right\} . \label{SWilson}
\end{equation}
The Grassmann-valued field $\xi \equiv \xi_{\alpha i} (x)$ has a spin index
$\alpha = 1, 2$ and a flavor index $i = 1, \ldots N$ that we leave implicit.
We denote by $\partial, \partial^{\ast}$, $\tilde{\partial}$ the forward,
backward and symmetric nearest neighbor differences on our cubic $T \times L$
lattice. The charge conjugation matrix $\mathcal{C}$ obeys
\begin{equation}
  \text{$\mathcal{C}$} \gamma_{\mu}  \text{$\mathcal{C}$}^{- 1} = -
  \gamma_{\mu}^{\top} = - \gamma_{\mu}^{\ast}, \quad \text{$\mathcal{C}$} = -
  \text{$\mathcal{C}$}^{\top} .
\end{equation}
For even $N$ each pair of Majorana fermions may be considered as one Dirac
fermion
\begin{equation}
  \psi = \frac{1}{\sqrt{2}} (\xi_1 + i \xi_2), \hspace{1em} \overline{\psi} =
  \frac{1}{\sqrt{2}} (\xi_1^{\top} - i \xi_2^{\top})\mathcal{C}
\end{equation}
with its independent $\psi, \overline{\psi}$. In the Majorana form the full
global symmetry group O($N$) is manifest beside (without Wilson term) the
discrete $\gamma_5$ symmetry whose spontaneous breaking was studied in
{\cite{Gross:1974jv}} in the $N \rightarrow \infty$ limit. The model is
renormalizable in the strict sense: there is no other O($N$) invariant scalar
4-fermion interaction term. For $N = 2$ \ we have the Thirring model
{\cite{Thirring:1958in}}, {\cite{Coleman:1974bu}}, the cases $N \geqslant 3$
are expected to be asymptotically free. In the remainder of this paper we set
the Wilson parameter to $r = 1$ and work in lattice units $a = 1$. The
discrete chiral symmetry $\xi \rightarrow \gamma_5 \xi$ of the massless
continuum theory is broken by the Wilson term and is only expected to be
recovered in the continuum limit at the critical mass \ $m = m_c$. On the
torus we consider four conceivable combinations of periodic or antiperiodic
boundary conditions in the two directions. Periodicity angles different from
$0, \pi$ {\emdash} as sometimes used for Dirac fermions \ {\emdash} would not
lead to a periodic action density for Majorana fermions. We label the possible
boundary conditions by a bit-vector
\begin{equation}
  \varepsilon_{\mu}, \quad \varepsilon_0, \varepsilon_1 \in \{0, 1\}
\end{equation}
where 0 stands for periodic and 1 for antiperiodic boundary conditions in the
corresponding direction.

Often the interaction term is factorized by the introduction of an auxiliary
bosonic field. For us it will be more convenient to think of $m \rightarrow m
(x)$ as an $x$-dependent mass for a while. If $Z_{\xi}^{\varepsilon} [m]$ is
the partition function of one free Majorana fermion with boundary condition
$\varepsilon_{\mu}$ in this background field, then the partition function of
the interacting theory is written as
\begin{equation}
  Z_{\tmop{int}} = \exp \left\{ \frac{g^2}{2} \sum_x
  \frac{\partial^2}{\partial m (x)^2} \right\} (Z_{\xi}^{\varepsilon} [m])^N .
  \label{Zint}
\end{equation}
Integrating the fermions in the remaining Gaussian problem yields the Pfaffian
\begin{equation}
  Z_{\xi}^{\varepsilon} [m] = \tmop{Pf} \left[ \mathcal{C}(\gamma_{\mu}
  \tilde{\partial}_{\mu} + m - \frac{1}{2} \partial^{\ast} \partial) \right] .
\end{equation}
In appendix~\ref{Pfaffapp} one can find a reminder of the definition of Pfaffians. For even $N$
we may replace the Pfaffians by the $N / 2$ power of the determinant. In this
form and with a factorizing field, the model can be simulated by standard
methods like HMC as carried out in {\cite{Korzec:2006hy}}. For larger $g$ this
compute-intensive task became rather difficult due to singularities developing
in the operator under the Pfaffian. Of course, the model itself remains
completely well-defined on any finite lattice. Fermionic and compact bosonic
variables are safe in this respect.

As a first step we now introduce the loop representation
{\cite{Gattringer:1998cd}} of $Z_{\xi}^{\varepsilon} [m]$. It may in fact also
be looked upon as a dimer ensemble similar to those derived in
{\cite{Rossi:1984cv}} for strong coupling QCD.

\section{External field fermion partition function}

As a building block for the Gross-Neveu models we consider for a single
Majorana field the external field action
\begin{equation}
  S = \frac{1}{2} \sum_x \varphi (x) \xi^{\top}_{} (x)\mathcal{C} \xi (x) -
  \sum_{x, \mu} \tau (x, \mu) \xi^{\top}_{} (x)\mathcal{C}P ( \hat{\mu}) \xi
  (x + \hat{\mu}) . \label{SFermi}
\end{equation}
We assume a lattice with $T$ sites in the time direction ($\mu = 0$) and $L$
sites in the space direction ($\mu = 1$). The variables $\varphi (x) = 2 + m
(x)$ and $\tau (x, \mu)$ are external commuting fields. The link field $\tau$
is introduced for completeness. It will be dropped at
some point. The lattice derivatives in (\ref{SWilson}) combine to Wilson
projectors that we define for arbitrary lattice unit vectors $n = \pm
\hat{\mu}$
\begin{equation}
  P (n) = \frac{1}{2} (1 - n_{\mu} \gamma_{\mu}), \quad ( \text{$\mathcal{C}$}
  P (n))^{\top} = - \text{$\mathcal{C}$} P (- n) .
\end{equation}
The last identity implies
\begin{equation}
  \xi^{\top}_{} (x)\mathcal{C}P ( \hat{\mu}) \xi (x + \hat{\mu}) =
  \xi^{\top}_{} (x + \hat{\mu})\mathcal{C}P (- \hat{\mu}) \xi (x) .
  \label{reverse}
\end{equation}
While the field $\xi$ has torus periodicity $\varepsilon$, the external fields
$\varphi, \tau$ are continued periodically to obtain a periodic action
density.

Defining the `covariant' projecting hop operator
\begin{equation}
  (H_{\mu} \xi) (x) = \tau (x, \mu)\mathcal{C}P ( \hat{\mu}) \xi (x +
  \hat{\mu})
\end{equation}
we may, with the help of (\ref{reverse}), also write the action in the
manifestly antisymmetric short-hand form
\begin{equation}
  S = \frac{1}{2} \sum_x \varphi \xi^{\top}_{} \mathcal{C} \xi - \frac{1}{2}
  \sum_{x, \mu} \xi^{\top}_{} (H_{\mu} - H_{\mu}^{\top}) \xi
\end{equation}
where $H_{\mu}^{\top}$ is transposed with respect to both spin and space
indices yielding
\begin{equation}
  (H_{\mu}^{\top} \xi) (x) = - \tau (x - \hat{\mu}, \mu)\mathcal{C}P (-
  \hat{\mu}) \xi (x - \hat{\mu}) .
\end{equation}
Now we are ready for the partition function
\begin{equation}
  Z_{\xi}^{\varepsilon} [\varphi, \tau] = \int D \xi \mathe^{- S} .
\end{equation}
The measure is
\begin{equation}
  D \xi = \prod_x (d \xi_1 d \xi_2) (x)
\end{equation}
and yields
\begin{equation}
  Z_{\xi} [\varphi, \tau] = \tmop{Pf} \left( \varphi \mathcal{C}- \sum_{\mu} (H_{\mu} -
  H_{\mu}^{\top}) \right)
\end{equation}
which is a nonlocal expression in the external fields as in the case of the
usual Dirac fermion determinant. In appendix~\ref{Pfaffapp} we evaluate the
Pfaffian for $\tau \equiv 1$ and constant $\varphi$ for all four boundary
conditions.

\section{Equivalent statistical systems}

\subsection{Dimer representation}

The Grassmannian Boltzmann factor may be expanded,
\begin{equation}
  Z_{\xi} = \int D \xi \prod_x \left\{ 1 + \varphi \xi_2 \xi_1 \right\} 
  \prod_{x, \mu} \left( 1 + \tau (x, \mu) \xi^{\top}_{} (x)\mathcal{C}P (
  \hat{\mu}) \xi (x + \hat{\mu}) \right) .
\end{equation}
All fields in the curly bracket are at $x$ and this factor is best considered
as part of the measure. We have here chosen $\mathcal{C}$ such that
$\frac{1}{2} \xi^{\top}_{} \mathcal{C} \xi = \xi_1 \xi_2$ which just amounts
to a phase convention. Note that the square of the hop-term vanishes due to
the one-dimensional projectors. There is only one linear combination of the
two Grassmann numbers contributing from each site which squares to zero. We
now introduce  one-bit-valued dimer or bond variables {\cite{Rossi:1984cv}}
on each link $k (x, \mu) = 0, 1$, whose values are used to organize the
expansion as
\begin{equation}
  Z_{\xi} [\varphi, \tau] = \sum_{\{k (x, \mu)\}} \int D \xi \prod_x \left\{ 1
  + \varphi \xi_2 \xi_1 \right\} \prod_{x, \mu} \left( \tau (x, \mu)
  \xi^{\top}_{} (x)\mathcal{C}P ( \hat{\mu}) \xi (x + \hat{\mu}) \right)^{k
  (x, \mu)} .
\end{equation}
As in {\cite{Rossi:1984cv}} the goal now is to integrate out the fermions to
yield a Boltzmann weight $\rho [k]$ for each dimer configuration. By asking
how the Grassmann integrations can be saturated site by site it is clear that
a non-zero weight only arises if at each site there are either two dimers
adjacent from different links or none at all. In the latter case the
integration is saturated by the measure term and a factor $\varphi (x)$
appears for this site. We also call these contributions monomers. Due to the
above constraint the dimers have to form closed non-intersecting and
non-backtracking loops. We choose randomly a starting point and an orientation
on such a loop such that along the loop one visits the sites $(x_1, x_2,
\ldots, x_l$). Consecutive sites differ by lattice unit vectors, $x_{i + 1} =
x_i + \hat{n}_i$, including $x_1 = x_l + \hat{n}_l$ in the last step. For such
a loop the product of bilinears
\begin{equation}
  (\xi^{\top}_{} (x_1)\mathcal{C}P ( \hat{n}_1) \xi (x_2)) \left(
  \xi^{\top}_{} (x_2)\mathcal{C}P ( \hat{n}_2) \xi (x_3) \right) \cdots \left(
  \xi^{\top}_{} (x_l)\mathcal{C}P ( \hat{n}_l) \xi (x_1) \right)
  \label{loopbi}
\end{equation}
has to be considered together with the integrations on the $l$ sites involved.
Note that here (\ref{reverse}) is relevant on the links that are transversed
in the negative direction. The trivial key formula is, for a single site,
\begin{equation}
  \int d \xi_1 d \xi_2 \xi \xi^{\top}_{} =\mathcal{C}^{- 1} .
\end{equation}
Now the expression (\ref{loopbi}) integrates to
\begin{equation}
  X = - \tmop{tr} [P ( \hat{n}_1) P ( \hat{n}_2) \cdots P ( \hat{n}_l)] .
  \label{protrace}
\end{equation}
Here appears the very important minus sign for a closed fermion loop,
well-known for instance from Feynman diagrams. It is not difficult to see that
this result is independent of the starting point and orientation chosen.

The evaluation of the spin factor follows {\cite{Stamatescu:1980br}}. Let us
introduce eigenspinors of the projectors
\begin{equation}
  \left. \left. P ( \hat{n}_i) = | \hat{n}_i \right\rangle \langle \hat{n}_i
  |, \quad \langle \hat{n}_i | \hat{n}_i \right\rangle = 1,
\end{equation}
\begin{equation}
  \left. \left. \left. X = - \langle \hat{n}_1 | \hat{n}_2 \right\rangle
  \langle \hat{n}_2 | \hat{n}_3 \right\rangle \cdots \langle \hat{n}_l |
  \hat{n}_1 \right\rangle
\end{equation}
A spinor is rotated by an angle $\theta$ by the unitary spin matrix $R
(\theta) = \exp [(\theta / 2) \gamma_0 \gamma_1]$. This allows us to write
\begin{equation}
  \left. \left. | \hat{n}_i \right\rangle = R (\Delta \theta_i) | \hat{n}_{i +
  1} \right\rangle \quad \Delta \theta_i \in \{0, \pm \pi / 2\}, \quad i = 1,
  2, \ldots, l \label{rotstep}
\end{equation}
with $\hat{n}_{l + 1} = \hat{n}_1$. Using
\begin{equation}
  \left. \langle \hat{n}_j | R (\Delta \theta_i) | \hat{n}_j \right\rangle =
  \cos (\Delta \theta_i / 2)
\end{equation}
we get to
\begin{equation}
  \left. X = - \langle \hat{n}_l | \hat{n}_1 \right\rangle \prod_{i = 1}^{l -
  1} \cos (\Delta \theta_i / 2) .
\end{equation}
The rotation accumulated in steps (\ref{rotstep}) is
\begin{equation}
  \left. \left. | \hat{n}_1 \right\rangle = R (\Theta - \Delta \theta_l) |
  \hat{n}_l \right\rangle  \quad \tmop{with} \quad \Theta = \sum_{i = 1}^l
  \Delta \theta_i .
\end{equation}
For closed paths we have $\Theta = 2 \pi \nu$ and $R (\Theta - \Delta
\theta_l) = \cos (\pi \nu) R (- \Delta \theta_l)$. For the nonzero lattice
angles, $\cos (\Delta \theta_i / 2) = 1 / \sqrt{2}$ . Altogether the final
result is
\begin{equation}
  X = (-)^{\nu + 1} 2^{- N_c / 2},
\end{equation}
where $N_c$ is the number of $\pm \pi / 2$ angles (`corners') occurring along
the loop and $\nu = 0, \pm 1$ is the number of complete rotations the loop
makes. The extra minus sign for $| \nu | = 1$ is the one associated with
fermions under $2 \pi$-rotations. If we include a non-trivial $\tau$ field
then the product of its Wilson loops for all dimer loops appear in addition in
the weight. After this remark we set $\tau \equiv 1$ until further notice.

Although we are on a lattice here, we can define homotopy classes of loops.
Two loops are homotopic to each other if they can be transformed into each
other by a sequence of steps, where dimers are only changed around a single
plaquette. We see from the above that $X$ is positive for all configurations
containing only loops that are homotopic to the trivial loop, just a point.
The two minus signs characteristic of fermions compensate each other in this
class!

\begin{figure}[htb] 
\centering
\epsfig{file=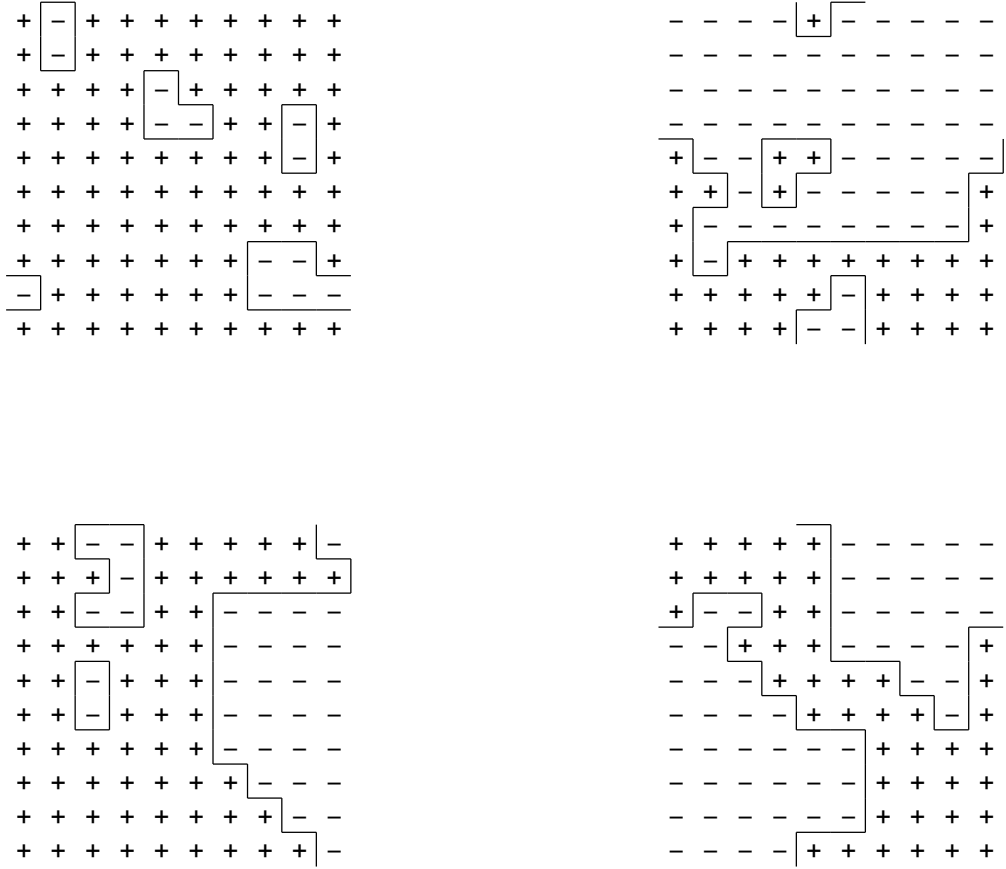,width=0.9\textwidth}
\caption{Examples of dimer configurations (solid lines) in the topological classes
$\mathcal{L}_{00}, \mathcal{L}_{10},\mathcal{L}_{01}, \mathcal{L}_{11}$ (from upper left to
lower right). The time direction ($\mu=0$) points to the right.
The signs refer to spins on the dual lattice.}
\label{classes}
\end{figure}

Loops can wind however around the torus in either direction as noted in
{\cite{Gattringer:2007em}}. A pair of loops winding around the same direction
is still in the trivial homotopy class. An odd number of windings leads
however to a new class. This may also happen in both directions at the same
time and hence there are the four classes $\mathcal{L}_{00}, \mathcal{L}_{10},
\mathcal{L}_{01}, \mathcal{L}_{11}$. In figure~\ref{classes} we show a
representative of each class. They are equilibrium configurations of free
fermions at $m = 0$ and $T = L = 10$. The meaning of the $+ / -$ signs in the
plots will become clear later. Only configurations from $\mathcal{L}_{00}$
have a positive weight while in the other cases there is an odd number of
closed loops with zero total rotation angle each of which contributes a factor
$- 1$. By introducing {\em{antiperiodic}} boundary conditions in some
direction the loops closing around that direction receive yet another sign
without changing the topologically trivial ones.

With the local weight
\begin{equation}
  \rho [k] = \prod_x f (k, x) \label{dimweight}
\end{equation}
with
\begin{equation}
  f (k, x) = \left\{ \begin{array}{ll}
    \varphi (x) & \tmop{if} \tmop{monomer} \tmop{at} x\\
    1 & \tmop{if} 2 \tmop{dimers} \tmop{in} \tmop{the} \tmop{same}
    \tmop{direction} \tmop{at} x\\
    1 / \sqrt{2} & \tmop{if} 2 \tmop{dimers} \tmop{in} \tmop{different}
    \tmop{directions} \tmop{at} x\\
    0 & \tmop{else}
  \end{array} \right.
\end{equation}
we now define the positive dimer partition functions
\begin{equation}
  Z_k^{00} [\varphi] = \sum_{\{k (x, \mu)\} \in \mathcal{L}_{00}} \rho [k],
  \quad Z_k^{10} (\varphi) = \sum_{\{k (x, \mu)\} \in \mathcal{L}_{10}} \rho
  [k], \quad \tmop{etc} .
\end{equation}
From what was said above the connections
\begin{eqnarray}
  4 Z_k^{00} [\varphi] & = & + Z_{\xi}^{00} [\varphi] + Z_{\xi}^{10} [\varphi]
  + Z_{\xi}^{01} [\varphi] + Z_{\xi}^{11} [\varphi] \\
  4 Z_k^{10} [\varphi] & = & - Z_{\xi}^{00} [\varphi] + Z_{\xi}^{10} [\varphi]
  - Z_{\xi}^{01} [\varphi] + Z_{\xi}^{11} [\varphi] \\
  4 Z_k^{01} [\varphi] & = & - Z_{\xi}^{00} [\varphi] - Z_{\xi}^{10} [\varphi]
  + Z_{\xi}^{01} [\varphi] + Z_{\xi}^{11} [\varphi] \\
  4 Z_k^{11} [\varphi] & = & - Z_{\xi}^{00} [\varphi] + Z_{\xi}^{10} [\varphi]
  + Z_{\xi}^{01} [\varphi] - Z_{\xi}^{11} [\varphi] 
\end{eqnarray}
arise, which can be inverted. If we want to realize the boundary conditions
$\epsilon_{\mu} = (1, 0)$ of {\cite{Korzec:2006hy}} (or actually any other
definite choice for the fermions) we have to sum over all dimer classes
including negative weight contributions
\begin{equation}
  Z_{\xi}^{10} [\varphi] = Z_k^{00} [\varphi] + Z_k^{10} [\varphi] - Z_k^{01}
  [\varphi] + Z_k^{11} [\varphi] . \label{Zap}
\end{equation}

All these relations between partition functions can be turned into relations
between expectation values of the scalar fermion density and monomer densities by
differentiating with respect to $\varphi (x)$. One example based on
(\ref{Zap}) is
\[ \begin{array}{ll}
     - \frac{\varphi (x)}{2} \left\langle \xi^{\top}_{} (x)\mathcal{C} \xi (x)
     \right\rangle_{10} & =
   \end{array} \]

\begin{equation}
  \frac{Z_k^{00} \left\langle K (x) \right\rangle_{00} + Z_k^{10} \left\langle
  K (x) \right\rangle_{10} - Z_k^{01} \left\langle K (x) \right\rangle_{01} +
  Z_k^{11} \left\langle K (x) \right\rangle_{11}}{Z_k^{00} + Z_k^{10} -
  Z_k^{01} + Z_k^{11}} . \label{Sap}
\end{equation}
The observable $K (x)$ is one if there is a monomer at $x$ and zero otherwise.
With the help of $\tau (x, \mu)$ as a source one could establish further
relations.

For free fermions ($\varphi = 2 + m$) in the thermodynamic limit at fixed $m >
0$ the various $Z_{\xi}^{\epsilon}$ differ only by \ exponentially small
amounts and $Z_k^{00}$ dominates among the dimer ensembles. Taking the
finite volume continuum limit ($L \rightarrow \infty$ with $L m$ fixed, see
appendix~\ref{Pfaffapp}) and in particular for $m = 0$ this is not so. In the
latter case we have an exact zero fermion mode for $\epsilon_{\mu} = (0, 0)$ and
$Z_{\xi}^{00} = 0$ holds. This implies
\begin{equation}
  Z_k^{00} = Z_k^{10} + Z_k^{01} + Z_k^{11} \hspace{1em} \tmop{for} m = 0.
\end{equation}

\subsection{Spin representation}

In this subsection we transform the dimer system to yet another representation
by Ising spins. This will allow us to design a global cluster algorithm. A
clue that this may be possible is given by the idea that a natural way to
manage and modify the closed loops in the dimer formulation is to consider
them as boundaries of domains of up-spins surrounded by down-spins (Peierls
contours).

\begin{figure}[htb]
\centering
  \resizebox{8cm}{!}{\epsfig{file=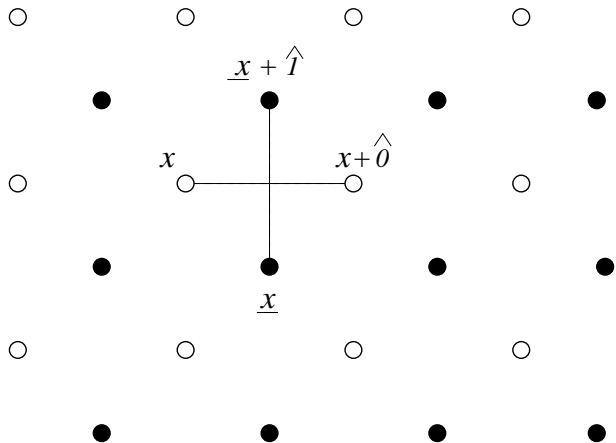}}
  \caption{Illustration of the labelling of the original
and the dual lattice.}
\label{dual}
\end{figure}

The spins that we introduce live on the lattice dual to the one carrying the
fermionic and the dimer variables. Its sites labelled by underlined
$\underline{x}$ are dual to plaquettes of the original lattice and are
imagined to be located at their centers. Analogously, the sites of the
original lattice are dual to plaquettes in the new one. Links $(x, \mu)$ and
$( \underline{x}, \underline{\mu}$) are dual to each other if they cross, see
figure~\ref{dual}. The idea is now to put an Ising field $s ( \underline{x})$ on
the dual lattice and to identify configurations
\begin{equation}
  k (x, \mu) = \left\{ \begin{array}{lll}
    1 & \tmop{if} & s ( \underline{x}) s ( \underline{x} +
    \widehat{\underline{\mu}}) = - 1\\
    0 & \tmop{if} & s ( \underline{x}) s ( \underline{x} +
    \widehat{\underline{\mu}}) = + 1
  \end{array} \right. . \label{ktos}
\end{equation}
In other words, dimers are located where nearest neighbor spins on the dual
lattice are antiparallel. In a first stage we restrict ourselves to the class
$\mathcal{L}_{00}$ of dimer configurations.

We first prove that for each admissible dimer configuration there are exactly
two spin fields obeying (\ref{ktos}) that differ by a global spin-flip. In a
first step we define a Z(2) lattice gauge field{\footnote{This field has
nothing to do with the earlier $\tau$.}} on the dual lattice in terms of the
dimers on the original lattice
\begin{equation}
  \sigma ( \underline{x}, \underline{\mu}) = \left\{ \begin{array}{lll}
    + 1 & \tmop{if} & k (x, \mu) = 0\\
    - 1 & \tmop{if} & k (x, \mu) = 1
  \end{array} \right. .
\end{equation}
Because of the constraints on $k$ this gauge field is unity when multiplied
around any plaquette (on the dual lattice). As we restrict ourselves to
$\mathcal{L}_{00}$ here, also loops around the torus are unity for this gauge
field. Thus $\sigma$ it a pure gauge on the torus with a periodic gauge
function. The spin-field is this gauge function that we construct now. We
choose a site $\underline{y}$ (the origin, for example) and set $s (
\underline{y}) = + 1.$ Now this value is parallel-transported with $\sigma$ to
all other sites, for instance along \ a maximal tree rooted at
$\underline{y}$. Due to the absence of curvature in $\sigma$, the result is
path-independent, consistent and unique. Starting from $s ( \underline{y}) = -
1$ we obtain the other configuration associated with $\{k (x, \mu)\}$. The
signs in figure~\ref{classes} are just these spins.

While we now have exactly two spin fields for each admissible dimer
configuration in $\mathcal{L}_{00}$, not all conceivable spin fields are
reached in this way. Obviously, spin configurations, that on any plaquette
look like
\begin{equation}
  \left(\begin{array}{cc}
    s_4 & s_3\\
    s_1 & s_2
  \end{array}\right) = \left(\begin{array}{cc}
    - & +\\
    + & -
  \end{array}\right), \left(\begin{array}{cc}
    + & -\\
    - & +
  \end{array}\right) \label{illegal}
\end{equation}
do not occur in the image, where we here just gave simple labels to the four
spins. They would correspond to crossing loops not allowed by the original
Grassmann variables. If this is however excluded on all plaquettes, then we can
reconstruct admissible $\{k (x, \mu)\}$ configurations \ from $\{s (
\underline{x})\}$. The total Boltzmann factor in the spin representation is
now a big product with one factor for each plaquette. These weights $w$ are
given in table~\ref{plweights} which lists only 8 of the 16 configurations and
is completed by using $w (s_1, s_2, s_3, s_4) = w (- s_1, - s_2, - s_3, -
s_4)$.
\begin{table}[htb]
\centering
    \begin{tabular}{|c|c|c|c|c|}
      \hline
      $s_1$ & $s_2$ & $s_3$ & $s_4$ & $w$\\
      \hline
      $+$ & $+$ & $+$ & $+$ & $\varphi (x)$\\
      \hline
      $+$ & $+$ & $+$ & $-$ & $1 / \sqrt{2}$\\
      \hline
      $+$ & $+$ & $-$ & $+$ & $1 / \sqrt{2}$\\
      \hline
      $+$ & $-$ & $+$ & $+$ & $1 / \sqrt{2}$\\
      \hline
      $-$ & $+$ & $+$ & $+$ & $1 / \sqrt{2}$\\
      \hline
      $+$ & $+$ & $-$ & $-$ & $1$\\
      \hline
      $+$ & $-$ & $-$ & $+$ & $1$\\
      \hline
      $+$ & $-$ & $+$ & $-$ & 0\\
      \hline
    \end{tabular}
\caption{Plaquette weights depending on the spin configuration.}
\label{plweights}
\end{table}
The monomer strength $\varphi (x)$ is taken here at the site $x$ of the
original lattice sitting at the center of the dual plaquette considered.

The derivation of this representation resembles the construction of the dual
formulation for generalized Ising models {\cite{Wegner:1984qt}}. We summarize
it (giving the ordinary self-dual two-dimensional Ising model as an example in
brackets): One first introduces new variables living on the bonds making up
the Hamiltonian of the original theory (a link field). Then the original spins
are summed over producing a constraint in the new variables (vanishing
plaquettes of the links interpreted on the dual lattice). This constraint is
then solved on the dual lattice (links given as a pure gauge by a site field).
The extension of the concept here is that we change from Grassmann
elements to bosonic variables, have an additional constraint (\ref{illegal})
to fulfill, and that there can be minus signs. One could talk of Fermi-Bose-
or super-duality.

The plaquette weight can also be written in terms of pairwise nearest neighbor
bond-interactions of the form [writing $\delta_{i j} \equiv \delta_{s_i s_j}
\equiv \frac{1}{2} (1 + s_i s_j)$]
\begin{eqnarray}
  w (s_1, s_2, s_3, s_4) & = & \text{}  \left\{ p [\delta_{12} + \delta_{23} +
  \delta_{34} + \delta_{41}] + q [\delta_{12} \delta_{34} + \delta_{14}
  \delta_{23}] + \right. \nonumber\\
  &  & \left. r [\delta_{12} \delta_{14} + \delta_{21} \delta_{23} +
  \delta_{32} \delta_{34} + \delta_{41} \delta_{43}] \right\},  \label{wbond}
\end{eqnarray}
with the $x$-dependence in $p, q, r$ suppressed. To match table~\ref{plweights},
the following equations have to hold
\begin{eqnarray*}
  \varphi (x) & = & 4 p + 2 q + 4 r\\
  \frac{1}{\sqrt{2} \text{}} & = & 2 p + r\\
  1 & = & 2 p + q.
\end{eqnarray*}
The solution of this system is
\begin{equation}
  r = \frac{1}{4} (\varphi - 2) = \frac{m}{4}, \quad p = \frac{1}{2 \sqrt{2}}
  - \frac{r}{2}, \quad q = 1 - \frac{1}{\sqrt{2}} + r. \label{bondprob}
\end{equation}
We remark here that all coefficients are positive for $0 \leqslant m \leqslant
2 \sqrt{2} \approx 2.83.$ In the free case, this clearly covers the relevant
range of bare masses.

If we now include dimer configurations in the other sectors by using
(\ref{ktos}), the only difference is that the resulting $s (x)$ is
antiperiodic in the direction {\em{orthogonal}} to those where dimer loops
run around the torus. Thus $\mathcal{L}_{10}$ corresponds to \ spin-fields
antiperiodic in {\em{space}}, $\mathcal{L}_{01}$ to those in {\em{time}}
and $\mathcal{L}_{11}$ to spins with both directions antiperiodic. The
mechanism here is that antiperiodic boundary conditions of the spins force an
interface into their configurations which leads to the nontrivial dimer loop
topology. Introducing further partition functions $Z_s$ for the \ spin
ensembles we relate
\begin{equation}
  Z_k^{00} = \frac{1}{2} Z_s^{00}, \hspace{1em} Z_k^{10} = \frac{1}{2}
  Z_s^{01}, \hspace{1em} Z_k^{01} = \frac{1}{2} Z_s^{10}, \hspace{1em}
  Z_k^{11} = \frac{1}{2} Z_s^{11} .
\end{equation}
The factor $1 / 2$ cancels the global spin-flip symmetry. Again, by
differentiation, we may relate expectation values, where the presence
(absence) of a monomer yielding $K (x) = 1 (0)$ in the spin language
translates into maximally `polarized' plaquettes where all four spins are
parallel,
\begin{equation}
  K (x) = \delta_{4, |M (x) |}, \hspace{1em} M (x) = \sum_{\underline{x}
  \tmop{around} x} s ( \underline{x}) .
\end{equation}

\section{Simulation algorithms}

\subsection{Local algorithms}

A local algorithm to simulate any one of the above dimer ensembles with all
weights taken positive was recently described and tested by Gattringer et al.
{\cite{Gattringer:2007em}}. The simplest case to consider is a free Majorana
fermion of mass $m$ with $\varphi (x) = 2 + m$. \ In the updates one actually
performs only changes that are local in the homotopy-sense by proposing
dimer-flips $k (x, \mu) \rightarrow 1 - k (x, \mu)$ around plaquettes. The
move is accepted with the Metropolis probability corresponding to the ratio of
$\rho$ in (\ref{dimweight}) for the new and the old configuration, which is a
locally computable quantity. Of course, it vanishes, if the new configuration
would violate one of the constraints. This update stays in the homotopy class
fixed by the starting configuration (see figure~\ref{classes}) and is ergodic within it. Thus
the various ensembles can be simulated which correspond to combinations of
periodic and antiperiodic Pfaffians. In \ {\cite{Gattringer:2007em}} it was
demonstrated that already such simulations are vastly more efficient than HMC
type simulations.

The entirely equivalent update in the Ising form consists of local spin-flips.
The Metropolis decision in this case depends on the eight nearest and
next-to-nearest (diagonal) neighbors that share plaquettes with the spin in
focus. The numerical efficiency in both forms is very similar.

\subsection{Cluster algorithm}

The plaquette interaction in (\ref{wbond}) is now written as a superposition
of 10 different terms, schematically given by
\begin{equation}
  \begin{array}{ll}
    w (s_1, s_2, s_3, s_4) & =
  \end{array} \sum_{i = 1}^{10} P_i \Delta_i (s_1, s_2, s_3, s_4)
\end{equation}
with $P_i \in \{p, q, r\}$ and $\Delta (s_1, s_2, s_3, s_4) \in \{0, 1\}$. In
complete analogy to {\cite{Swendsen:1987ce}} we introduce now ten-valued bond
variables $b (x)$ with each value corresponding to one of these
terms to obtain
\begin{equation}
  Z_s = \sum_{\{ \text{$b (x)$, $s ( \underline{x})$} \}} 
  \prod_{\tmop{plaq}} P_{b (x)} \Delta_{\text{$b (
  x)$}} (s_1, s_2, s_3, s_4) .
\end{equation}
To avoid a too clumsy notation here again $s_1, s_2, s_3, s_4$ are spins
around each of the plaquettes. The celebrated trick of
{\cite{Swendsen:1987ce}} consists of Monte Carlo sampling {\em{both}} the
$b$ and the $s$ variables. As the first part of an update cycle one chooses
new $b$ at fixed $s$ by a local heatbath procedure. Of course, in general, the
choice is between less than 10 possibilities, as some of the $\Delta_{\text{$b
(x)$}} (s_1, s_2, s_3, s_4)$ vanish. Then, for given bonds $b$,
any of the $2^{T L}$ spin configurations has either weight zero or a constant
nonzero weight \ depending on the constraints given by the product of all
factors $\Delta_{\text{$b (x)$}}$. Just as in the standard Ising
model case we now construct the percolation clusters defined by the active
bonds in all $\{\Delta_{\text{$b (x)$}} \}$. By flipping the
spins in each cluster as a whole with probability 1/2, we sample one of the
allowed and equally weighted spin configurations. The overall procedure
amounts to a global independent sampling of spins (at fixed bonds) \ and will
be numerically demonstrated to almost eliminate critical slowing in section
\ref{numtest}.

At the stage of selecting a new spin field one may also construct improved
cluster estimators. This is achieved, if, for some observable, one is able to
analytically average over all conceivable spin-assignments of which only one
is taken as the next configuration.

While the above procedure is analogous to the well-known Swendsen-Wang
algorithm {\cite{Swendsen:1987ce}} we could also study a single cluster
variant {\cite{Wolff:1988uh}}: one spin is chosen at random, and then only the
one cluster connected to it is constructed by investigating the plaquette
terms touched in the growth process until it stops. Then spins on this cluster
are always flipped. This may well be even more efficient as large clusters are
preferred.

We end this subsection with a remark on the global spin flip symmetry. At
first one may think that it is a (slightly) annoying redundancy in the new
representation. However, it is in fact essential to be able to grow clusters
whose energy (action) is associated with the surface (in our case the loops)
and not with the bulk. Loosely speaking, as one grows a (single) cluster,
there is always an energetic `way out' by flipping the whole lattice. Of
course, if this is all that happens, that algorithm will not be efficient. The
auxiliary percolation problem allows to find nontrivial clusters.

\subsection{Fluctuating boundary conditions}

In (\ref{Zap}) we saw that it is desirable to also be able to simulate
enlarged ensembles where one sums beside configurations of spins also over
several possible boundary conditions. If in conventional simulations one
proposes a change of the boundary conditions at fixed spins, one generates
energy (action) proportional to $T$ or $L$ and the proposal will practically
always be rejected. It was noticed however in {\cite{Hasenbusch:1992eu}} that
with cluster algorithms the situation can be different. In the step where we
pick new spins at fixed bonds the search among possible equally weighted new
configurations can be enlarged to also include changed boundary conditions. If
we label them by $\varepsilon_{\mu}$ again (for the spins now) the four
possible $\varepsilon$ become a dynamical variable. In
{\cite{Hasenbusch:1992eu}} these changes were introduced in a
single-cluster/Metropolis spirit, which would also be possible {\emdash} in
fact less involved {\emdash} here. In view of the future construction of
improved estimators we stick however to the many-cluster view and design now a
correspondingly generalized cluster algorithm.
It consists of the following steps:
\begin{itemize}
  \item We throw bonds on the links as discussed before.
  
  \item We determine by some percolation algorithm (e.g. tree search) the
  independent spin clusters connected by bonds {\em{but ignore}} two layers
  of links such that the torus is cut open. We take $\{(x, 0) |_{x_0 = T - 1}
  \}$ and $\{(x, 1) |_{x_1 = L - 1} \}$. We call these clusters preclusters,
  their connectivity is determined in the `interior'. Each of them carries a
  unique cluster label as a result.
  
  \item Now the remaining links are examined as far as they have been
  activated. We call these bonds clamps. They have the effect of sewing up
  (some of) the preclusters. This is done by the pointer technique described
  in {\cite{PhysRevB.14.3438}}. We may visualize the process as a graph with
  the preclusters as blobs, some of which get connected by lines.
  
  \item In this process \`{a} la {\cite{PhysRevB.14.3438}} one can detect when
  closed loops in the graph are formed. We set one of four types of flags
  whenever a loop is closed. They distinguish whether an odd or an even number
  of temporal or spatial clamps are met around the loop. We end with flags
  $f_{00}, f_{10}, f_{01}, f_{11}$ each being zero or one.
  
  \item Boundary conditions $\varepsilon$ are only compatible with this loop
  structure if for all occurring loops with flag up ($f_{\alpha \beta} = 1$)
  the condition $\alpha \varepsilon_0 + \beta \varepsilon_1 = 0 \tmop{mod} 2$
  holds. Of course, $\varepsilon = (0, 0)$ is always allowed.
  
  \item Among the compatible $\varepsilon$ (1,2 or 4 values) one is chosen
  with equal probability.
  
  \item Now flips for all preclusters are determined and executed. Connected \
  components of the graph flip together, but preclusters within these
  components can flip relative to each other if the boundary condition has
  changed. The above construction guarantees that the orientations thus
  propagated do not depend on the path that is taken on the graph.
\end{itemize}
Although rather short and compact in the end this is not a trivial code to
write. It is helpful to organize it under a geometric point of view focussing
on parallel transport between preclusters with a Z(2) group, where the
boundary conditions are gauge variables on the clamps. Of course, as we can
solve the free fermion ensembles exactly (appendix~\ref{Pfaffapp}) and easily get
high accuracy with cluster simulations, many significant checks by short
simulations (taking seconds) were available. A very good monitor for debugging
at every stage is to set traps for the occurrence of illegal plaquettes
(\ref{illegal}). An alternative strategy to move boundary conditions would be
to turn the loop structure of the above graph into a system of linear
equations in the Galois field of two elements (addition isomorphic to logical
{\tt xor}). Following {\cite{Brower:1990ng}} and {\cite{Bunk:1991zy}}
this can be solved by Gauss elimination.

The above scheme contains some nested pointer operations. One could be worried
in principle whether the execution time grows more than linearly with the
lattice volume. In practice there was found to be absolutely no problem of
this kind. This is in fact the same for cluster simulations of the standard
Ising model using the algorithm of {\cite{PhysRevB.14.3438}}. In our case the
problem is even less severe as the number of clamps is smaller than $T + L$,
not proportional to the volume.

\subsection{Negative mass}

For free fermions one could content oneself with the parameter range $m
\geqslant 0 = m_c$. On the other hand all results in appendix~\ref{Pfaffapp}
can be taken at arbitrary $m$. After all, the partition function on the finite
lattice is just a polynomial. When we later come back to the interacting
theory, it will also turn out, that negative mass fermions will be required
because $m_c < 0$ due to renormalization. The local algorithm
{\cite{Gattringer:2007em}} works for negative $m$, a sign problem only arises
if $\varphi = 2 + m$ changes sign. The bond probabilities (\ref{bondprob})
however restrict the cluster algorithm so far to $m \geqslant 0$. Luckily,
there is an alternative decomposition of the plaquette interaction into bonds
that comes to rescue when $m (x)$ is negative.

The $r$-term in (\ref{wbond}) is replaced by
\[ \frac{\tilde{r}}{2} [\delta_{12} \overline{\delta}_{14} + \delta_{21}
   \overline{\delta}_{23} + \delta_{32} \overline{\delta}_{34} + \delta_{41}
   \overline{\delta}_{43} + \overline{\delta}_{12} \delta_{14} +
   \overline{\delta}_{21} \delta_{23} + \overline{\delta}_{32} \delta_{34} +
   \overline{\delta}_{41} \delta_{43}], \label{wbondneg} \]
where we introduced also antibonds $\overline{\delta}_{i j} \equiv 1 -
\delta_{i j}$. The two other terms are unchanged but the coefficients are now
$\tilde{p}, \tilde{q}$. Using $\delta_{12} \overline{\delta}_{14} =
\delta_{12} - \delta_{12} \delta_{14}$ etc. one sees that the new weight
coincides with the old one if we identify $r = - \tilde{r}$ and $p = \tilde{p}
+ \tilde{r}$. The matching equations are now
\begin{eqnarray*}
  \varphi (x) & = & 4 \tilde{p} + 2 \tilde{q}\\
  \frac{1}{\sqrt{2}} & = & 2 \tilde{p} + \tilde{r}\\
  1 & = & 2 \tilde{p} + \tilde{q} + 2 \tilde{r},
\end{eqnarray*}
solved by
\begin{equation}
  \tilde{r} = \frac{1}{4} (2 - \varphi) = \frac{- m}{4}, \quad \tilde{p} =
  \frac{1}{2 \sqrt{2}} - \frac{\tilde{r}}{2}, \quad \tilde{q} = 1 -
  \frac{1}{\sqrt{2}} - \tilde{r} . \label{bondprobneg}
\end{equation}
Now all weights are positive for $- m \leqslant 2 (2 - \sqrt{2}) \approx
1.17$.

The presence of antibonds is compatible with the cluster search including
fluctuating boundary conditions. With an $m (x)$ that changes sign over one lattice, one
actually decomposes some plaquettes with (\ref{bondprob}) and others with
(\ref{bondprobneg}). One must however not make the mistake to think of the
preclusters as ferromagnetic (Weiss) domains, they contain in general both up
and down spins. This is why we took care to talk about distributing flips to
clusters rather than assigning new spin orientations to them as whole.

\section{Numerical applications\label{numtest}}

We now report on numerical experiments. In this first publication on the
method we stick to free fermions and the observable $K (x)$. It corresponds to
the scalar fermion density \ and is mainly used to diagnose the algorithm.
Hence, with the results of appendix~\ref{Pfaffapp}, every computed mean value
is known exactly and was verified to be reproduced by the Monte Carlo
simulations within errors. We do not plot results for $K$. They are too
boring: errors not visible on the graph and exact results agreeing with the
data within 1 and occasionally up to around 2 sigma.

For the algorithm, the non-interacting case does not seem to be fundamentally
different from the interacting Gross-Neveu model. All details necessary for
this extension are given, but the numerical implementation is deferred to a
future investigation. What remains to be seen is how the correlation between
monomers of different flavor influences the Monte Carlo dynamics at stronger
coupling.

\subsection{Critical slowing}

We performed a series of simulations of one species of free Majorana fermions
at the critical value $m = 0$. In this case, the only infrared scale is given
by the system size $T = L = 8 \ldots 128$. We simulated the trivial ensembles
corresponding to the loop class $\mathcal{L}_{00}$. Results are summarized in
table~\ref{csdtab}. Each run with the local algorithm, passing through the
lattice in lexicographic order, consists of $10^6$ sweeps of which a small
fraction{\footnote{We discard the first $(T / 16)^2 \times 1000$ sweeps in the local runs.}} is
discarded for thermalization. The autocorrelation time $\tau_{\tmop{int}, K}$
has been defined and measured as described in {\cite{Wolff:2003sm}}.

\begin{table}[htb] 
\centering
  \begin{tabular}{|c|c|c|l|l|c|}
    \hline
    & exact &\multicolumn{2}{c|} {local} & \multicolumn{2}{c|} {cluster} \\ 
    \hline
    $L$ & $K$ & $K$ & \multicolumn{1}{c|} {$\tau_{\tmop{int}, K}$} & 
    \multicolumn{1}{c|} {$K$} & $\tau_{\tmop{int}, K}$\\
    \hline
    8 & 0.80738 & 0.8080(5) & \phantom{44}5.13(7) & 0.80774(32) & 1.55(2)\\
    \hline
    16 & 0.78914 & 0.7885(4) & \phantom{4}14.4(3) & 0.78932(21) & 1.90(2)\\
    \hline
    32 & 0.77951 & 0.7795(4) & \phantom{4}44.0(1.6) & 0.77949(13) & 2.29(3)\\
    \hline
    64 & 0.77466 & 0.7742(5) & 187(17) & 0.77464(8) & 2.84(4)\\
    \hline
    128 & 0.77223 & 0.7721(4) & 444(58) & 0.77221(5) & 3.50(5)\\
    \hline
  \end{tabular}
  \caption{Monomer density and its integrated autocorrelation time
for local and cluster simulations at $m=0$ and lattice sizes $T=L$.}
\label{csdtab}
\end{table}

As one expects for local algorithms we see a steeply rising autocorrelation
time hinting at a dynamical exponent not too far from two --- we have no
ambition here to determine it precisely which would be very costly. One
notices, that the error (at fixed sweep number) is almost independent of $L$.
This means the variance just compensates the growing autocorrelation time and
decays roughly proportionally to $1 / T L$. This is in fact implied by scaling
and the canonical dimension of the (connected) 2-point function of the scalar
density. Although the integral over the autocorrelation function at $L = 128$
does not look too unconvincing, one may suspect that our number for
$\tau_{\tmop{int}, K}$ may only be a lower bound for this case.

The cluster simulations in the last columns consist of $0.6 \times 10^6$
sweeps which in our implementation takes about the same time as the local runs
on a single PC. We see small slowly rising autocorrelation times. From the two
largest lattices one would estimate an effective dynamical exponent
$z_{\tmop{eff}} \approx 0.30$, which is a typical value for cluster
algorithms. In total only about 15 CPU hours went into these demonstrations.
All codes have been programmed in {\tt MATLAB} and the update routine has
about 100 lines (50 without fluctuating boundary conditions).

The cluster algorithm performs very similarly also in the other topological
sectors. We did the $T = L = 128$ runs with spin boundary conditions
$\varepsilon_{\mu} = (0, 1)$ ($\mathcal{L}_{10})$ and found $\tau_{\tmop{int},
K} = 2.68 (4)$ and $\varepsilon_{\mu} = (1, 1)$ ($\mathcal{L}_{11})$ with \
$\tau_{\tmop{int}, K} = 2.29 (3)$, even shorter than the trivial sector.

\begin{figure}[htb] 
  \centering
  \epsfig{file=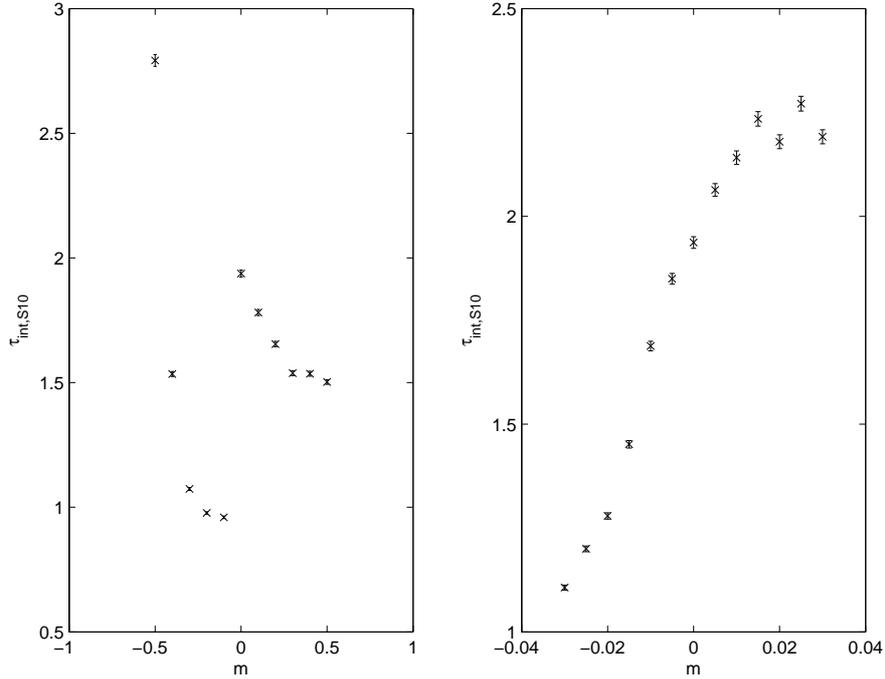,width=0.8\textwidth}
  \caption{Integrated autocorrelation time for the scalar density (\ref{S10})
  corresponding to fermion boundary conditions antiperiodic in one direction
  and periodic in the other at $T=L=128$. }
\label{taufig}
\end{figure}

The next series of runs to be reported is on $T = L = 128$ lattices at several
positive and negative masses. Again each data point is produced by $0.6 \times
10^6$ sweeps. These simulations included fluctuating boundary conditions.
Recorded observables were the monomer density $K$, the boundary conditions
$\varepsilon_{\mu}$ and the topological flags $f_{\alpha \beta}$. From these
data the distribution of  boundary conditions can be deduced and we
checked their correctness. As an example we consider here (\ref{Sap}) and
compute for the right hand side
\begin{equation}
  S_{10} = - \frac{2 + m}{2} \left\langle \xi^{\top}_{} (x)\mathcal{C} \xi (x)
  \right\rangle_{10} = \frac{\left\langle K \right\rangle - 2 \left\langle K
  \delta_{\varepsilon, (1, 0)} \right\rangle}{1 - 2 \left\langle
  \delta_{\varepsilon, (1, 0)} \right\rangle}, \label{S10}
\end{equation}
where on the right hand side the Ising ensemble with fluctuating boundary
conditions $\varepsilon$ is meant. The result at $m = 0$ is $S_{10} = 0.76987
(6)$ with the exact value being 0.769800361... Errors for this combination
of observables are estimated as discussed in {\cite{Wolff:2003sm}}, where the
definition of
$\tau_{\tmop{int}, S_{10}}$ from the fluctuations relevant for this
quantity can be found. We show these autocorrelation times in figure~\ref{taufig}. There seems
to be a steep rise by about one unit close to $m = 0$. The second plot shows a
better resolution of its vicinity. All these numbers stay comfortably small.
The combination measured here has no serious sign problem. One could however
construct positive cluster estimators for numerator and denominator. 

A simple example for the use of an improved estimator exploiting the
topological information can be given by the two observables with equal mean
\begin{equation}
  \left\langle \delta_{\varepsilon, (1, 0)} \right\rangle = \left\langle
  \frac{1}{4} \delta_{(f_{10}, f_{01}, f_{11}), (0, 0, 0)} + \frac{1}{2}
  \delta_{(f_{10}, f_{01}, f_{11}), (0, 1, 0)} \right\rangle .
\end{equation}
The fractions on the right hand are the part of admissible boundary
conditions given by $\varepsilon = (1, 0)$. Flag
positions not tagged here do not allow this value at all.
At $m = 0$ we find in our
run for the left hand side $0.18650 ($66) [$\tau_{\tmop{int}} = 0.852 (6)$]
and for the other mean $0.18650 ($37) [$\tau_{\tmop{int}} = 1.38 (1)$] while
the exact answer is 0.186455866.... Of course, the two estimates are strongly
correlated. It is clear that the left estimate is `more stochastic' using the
actually picked boundary conditions in the run and $\tau_{\tmop{int}}$ is
hence smaller. This pattern is typical for cluster estimators with the reduced
variance usually overcompensating this effect.

\subsection{Four fermion interaction}

To simulate the interacting Gross-Neveu model (\ref{SWilson}) we represent
each factor $Z_{\xi}^{\varepsilon}$ in (\ref{Zint}) by spin-ensembles. Their
$\varepsilon$ are summed over independently with weights depending on the
desired boundary conditions for the fermions. At each site $x$ an overall
configuration holds $0 \leqslant n (x) \leqslant N$ monomers. The only thing
that the interaction does is to change the corresponding weight into a factor
\begin{equation}
  \exp \left\{ \frac{g^2}{2}  \frac{\partial^2}{\partial m^2} \right\} (2 +
  m)^n = \sum_{j = 0}^{[n / 2]}  \frac{n!}{2^j j! (n - 2 j) !} g^{2 j} (2 +
  m)^{n - 2 j} = c (n, m, g) .
\end{equation}
We update one of the spin flavors at a time. Let us introduce the number
$\overline{n} (x)$ of monomers in the other momentarily frozen spins. Then in
our update the effective monomer weight (or local mass) has to be taken as
\begin{equation}
  \varphi |_{\overline{n}} (x) = \frac{c ( \overline{n} (x) + 1, m, g)}{c (
  \overline{n} (x), m, g} .
\end{equation}
The first few terms are
\begin{equation}
  \varphi |_0 = 2 + m, \quad \varphi |_1 = 2 + m + \frac{g^2}{2 + m}, \quad
  \varphi |_2 = 2 + m + 2 g^2  \frac{2 + m}{(2 + m) + g^2}, \ldots .
\end{equation}
In interacting theories the mass of Wilson fermions undergoes additive
renormalization. Both in the Thirring model ($N = 2)$ and in the higher $N$
Gross-Neveu model perturbative and nonperturbative calculations
{\cite{Korzec:2006hy}} as well as large $N$ approximations yield negative
values for the critical mass $m_c (g^2)$ close to which one wants to simulate.
This is why it was crucial to extend the cluster algorithm to also accommodate
$\varphi < 2$.

\subsection{Additional gauge coupling}

An amusing exercise is to add to the Majorana `flavor' group a `color' Z(2)
gauge interaction. We describe it here only very briefly. We now make use of
the gauge links in (\ref{SFermi}) and consider
\begin{equation}
  Z_{\tmop{color}} = \sum_{\tau (x, \mu)} \mathe^{\beta \sum_p \tau_p}
  (Z_s^{\varepsilon} [m, \tau])^N
\end{equation}
with the plaquette field $\tau_p$ made from the links $\tau (x, \mu) = \pm 1$.
The self-interaction can always be added, changing only the monomer weights.
We suppress it here. Each dimer loop receives as an additional factor the
Wilson loop made of $\tau$. By Stokes' theorem they can be replaced by a
product of $\tau_p$ either on all plaquettes where a $+$ spin resides or on
those where the negative spins sit, see figure~\ref{classes}. Let us choose $+$.
In the case of antiperiodic boundary conditions additional loops around the
torus appear where the torus closes (where the clamps were). As a
two-dimensional gauge theory is rather trivial the $\tau$ sum can be carried
out. We need to know how many plaquettes get tiled with $\tau_p$ an odd number
of times. Let us introduce the `composite' flavor spin-field
\begin{equation}
  S ( \underline{x}) = s^{(1)} ( \underline{x}) s^{(2)} ( \underline{x})
  \cdots s^{(N)} ( \underline{x}) .
\end{equation}
Then
\begin{equation}
  M_{\pm} = \sum_{\underline{x}} \delta_{S ( \underline{x}), \pm}
\end{equation}
contain this information and the extra weight from summing over $\tau$ is
\[ 2^{2T V} [\cosh (\beta)^{M_+} \sinh (\beta)^{M_-} + \cosh (\beta)^{M_-}
   \sinh (\beta)^{M_+}] . \]
This is equivalent to a magnetic field
\begin{equation}
  H = - \frac{1}{2} \ln \tanh (\beta)
\end{equation}
coupled to $S$ and fluctuating in sign. In addition we only get contributions
if an even number of flavors is antiperiodic in each direction separately (due
to Z(2) confinement). The fluctuating magnetic field can presumably be
included in the cluster simulation without problems. As we update a given
flavor, each spin can in addition bond to one exterior (`phantom') spin.

\section{Conclusion and Outlook}

It seems that with the new cluster algorithm the Gross-Neveu model (and the
Thirring model) are wide open for high precision simulation. Further
observables, in particular cluster estimators, remain to be constructed. The
O($N$) Noether currents may be accessible, for example. Also ratios of
partition functions like $Z_s^{00} / Z_s^{10}$ etc. are expected to have a
continuum limit and may serve as renormalized finite volume couplings. 
Majorana fermions are
prominent in supersymmetry. Maybe some studies on two dimensional
supersymmetry become possible with simulations very close to the continuum
limit.

An obvious question that every reader will have is if any of this carries over
to higher dimension and/or more complicated gauge interactions. Let us first caution here:
fermions in one space dimension are very special. This has manifested itself
in other so-called Fermi-Bose equivalences in two dimensions. At the heart of
this in operator language is the fact that the Jordan-Wigner
{\cite{Jordan:1928wi}} transformation transforms anticommuting to commuting
degrees of freedom {\em{without}} generating non-localities. This has no
obvious generalization to higher dimension (see however
{\cite{Fradkin:1980gt}}). The fact that we find positive weights for all
topologically trivial loops in a way seems to be the euclidean
counterpart. Also that Majorana fermions are in some sense equivalent to Ising
spins it not new, of course, see {\cite{Schultz:1964fv}},
{\cite{Kogut:1979wt}}. The main achievement here is that we simulate in a
standard (lattice) euclidean fermion formulation and can get really critical.
In higher dimension a dimer representation for fermions can probably be
constructed along similar lines, but the weight will be sign-fluctuating in a
more essential fashion. The slight hope may be that one could be able to
handle this sign problem with cluster estimators. In addition, the coupling to
gauge fields contributing fluctuating Wilson loops is an open problem. There
are ongoing efforts to simulate discrete models dual to nonabelian gauge
theories (spin foam) (see {\cite{Cherrington:2007ax}} and references therein).
This may be an interesting view on gauge theories in the context of the
approach to fermions developed here. In any case, the goal is attractive
enough to warrant further thought.

We would like to acknowledge discussions about the Gross-Neveu model
on the lattice with
Francesco Knechtli,
Bj\"{o}rn Leder, Rainer Sommer and most of all Tomasz Korzec.
We also thank the
Deutsche Forschungsgemeinschaft (DFG)
for support
in the framework of SFB Transregio~9.

\begin{appendix}
\section{Exact Majorana partition functions\label{Pfaffapp}}

The Pfaffian is defined for an antisymmetric matrix $A_{i j}$ of even size,
$i, j = 1, \ldots, 2 n$,
\begin{equation}
  \tmop{Pf} (A) = \int d^{2 n} \xi \mathe^{- \frac{1}{2} \xi^{\top} A \xi} =
  \frac{1}{2^n}  \frac{1}{n!} \epsilon_{i_1 i_2 \ldots i_{2 n}} A_{i_1 i_2}
  A_{i_3 i_4} \cdots A_{i_{2 n - 1} i_{2 n}}, \label{Pfdef}
\end{equation}
where we integrate over $2 n$ Grassmann variables and a sign convention for
the measure is implied. By a change of variables $\xi \rightarrow F \xi$ one
sees the well known identity
\begin{equation}
  \tmop{Pf} (F^{\top} A F) = \det (F) \tmop{Pf} (A), \label{Pftrans}
\end{equation}
and by squaring the integral in (\ref{Pfdef}),
\begin{equation}
  [\tmop{Pf} (A)]^2 = \det (A)
\end{equation}
follows.

We are interested in $A = \mathcal{C}(\gamma_{\mu} \tilde{\partial}_{\mu} + m
- \frac{1}{2} \partial^{\ast} \partial)$ with boundary conditions
$\varepsilon$. The determinant immediately follows by Fourier diagonalization
\begin{equation}
  \det_{\varepsilon} (A) = \prod_p ( \mathring{p}^2 + M (p)^2)
\end{equation}
with
\begin{equation}
  p = \left( \frac{2 \pi}{T} (n_0 + \varepsilon_0 / 2), \frac{2 \pi}{L} (n_1 +
  \varepsilon_1 / 2) \right), \hspace{1em} n_0 = 0, \ldots, T - 1,
  \hspace{1em} n_1 = 0, \ldots, L - 1
\end{equation}
and
\begin{equation}
  \mathring{p}_{\mu} = \sin (p_{\mu}), \hspace{1em} M (p) = m + \frac{1}{2}
  \hat{p}^2, \hspace{1em} \hat{p}_{\mu} = 2 \sin (p_{\mu} / 2) .
\end{equation}
In this article we also want to know the relative signs of $\tmop{Pf} (A)$ for
different boundary conditions and possibly negative $m$. This information is
lost in the determinant and we proceed differently.

We define the unitary Fourier transformation matrix
\begin{equation}
  F_{x p} = \frac{1}{\sqrt{T L}} \; \mathe^{i p x}
\end{equation}
which is augmented trivially to include a unit matrix in spin space. Then,
using (\ref{Pftrans}) we get
\begin{equation}
  \det (F) \tmop{Pf} (A) = \tmop{Pf} ( \tilde{A}) \text{}
\end{equation}
with
\begin{equation}
  \widetilde{A_{}}_{q p} =  \text{$\mathcal{C}(\gamma_{\mu} \mathring{p}_{\mu} + M
  (p))$} \delta_{p + q, 0} .
\end{equation}
The factor $\det (F)$ is just a phase that we fix later.

The matrix $\tilde{A}$ consists of antisymmetric blocks where momenta $p, - p$
get paired. If they are different (modulo $2 \pi$) such a block contributes
{\em{one}} factor $( \mathring{p}^2 + M (p)^2)$ to the Pfaffian.

For simplicity we now restrict our discussion to both $T$ and $L$ even. 
Then
all momenta get paired non-trivially except $p = (0, 0),
(\pi, 0), (0, \pi), (\pi, \pi)$
in the all periodic case $\varepsilon_{\mu} = (0, 0)$.
 We may summarize this
result by
\begin{equation}
  \tmop{Pf}_{\varepsilon} (A) = \prod_p \sqrt{\mathring{p}^2 + M (p)^2} \times
  \left\{ \begin{array}{ll}
    \tmop{sign} [m (m + 4)] \tmop{for} \varepsilon_{\mu} = (0, 0) & \\
    1 \hspace{26mm} \tmop{other} \varepsilon_{\mu} & 
  \end{array} \right. .
\end{equation}
Possible extra phase factors can be excluded by considering the limit $m
\rightarrow \infty$. Note that the double periodic Pfaffian changes sign at $m
= 0$.

In this article we need for comparison the scalar condensate in ensembles that
in fermion language have the partition function
\begin{equation}
  \overline{Z}_{\xi} [m] = \sum_{\varepsilon} c (\varepsilon)
  Z_{\xi}^{\varepsilon} [m] .
\end{equation}
It is given by
\begin{equation}
  - \frac{1}{2} \left\langle \xi^{\top}_{} \mathcal{C} \xi \right\rangle_c =
  \frac{1}{T L} \frac{\partial}{\partial m} \ln ( \overline{Z}_{\xi}) = -
  \frac{1}{2} \frac{\sum_{\varepsilon} c (\varepsilon) z (\varepsilon, m)
  \left\langle \xi^{\top}_{} \mathcal{C} \xi
  \right\rangle_{\varepsilon}}{\sum_{\varepsilon} c (\varepsilon) z
  (\varepsilon, m)}
\end{equation}
with
\begin{equation}
  z (\varepsilon, m) = \frac{Z_{\xi}^{\varepsilon} [m]}{\overline{}
  \sum_{\varepsilon'} Z_{\xi}^{\varepsilon'} [m]}
\end{equation}
and
\begin{equation}
  - \frac{1}{2} \left\langle \xi^{\top}_{} \mathcal{C} \xi
  \right\rangle_{\varepsilon} = \frac{1}{T L} \sum_p \frac{M (p)}{\mathring{p}^{2}
  + M^2 (p)} .
\end{equation}
All these exact results can be easily evaluated for any finite lattice
that is simulated. For $\varepsilon_{\mu} = (0, 0), m \to 0$ the
product $z \left\langle \xi^{\top}_{} \mathcal{C} \xi \right\rangle_{(0, 0)}$
remains finite but requires precaution numerically.

The continuum limit of $z$ can presumably be computed analytically. We here
content ourselves with the numerical construction of their Symanzik expansion
in a few cases. We set $T = L$ (aspect ratio one) and
\begin{equation}
  z (\varepsilon, m) |_{m L = \kappa} \simeq \sum_{k \ge 0} d_k (\kappa,
  \varepsilon) L^{- k}
\end{equation}
and compile some values in table~\ref{ztab}.

\begin{table}[htb] 
\centering
  \begin{tabular}{|c|c|l|l|l|}
    \hline
    $\kappa$ & $\varepsilon$ & $d_0$ & $d_2$ & $d_4$\\
    \hline
    0 & (1,0) & \phantom{-}0.3135575596 & -0.219897 & -1.1\\
    \hline
    $\kappa$ &  & $d_0$ & $d_1$ & $d_2$\\
    \hline
    1 & (0,0) & \phantom{-}0.12681663 & -0.048092 & -0.108\\
    \hline
    1 & (1,0) & \phantom{-}0.27632955 & \phantom{-}0.012935 & -0.135\\
    \hline
    -1 & (0,0) & -0.16991195 & -0.08631 & \phantom{-}0.183\\
    \hline
    -1 & (1,0) & \phantom{-}0.37023294 & \phantom{-}0.030368 & -0.285\\
    \hline
  \end{tabular}
  \caption{Symanzik expansion coefficients for ratios of partition functions at different
boundary conditions in the finite volume continuum limit at fixed $\kappa=mL$.}
\label{ztab}
\end{table}

The analysis of the asymptotic series was carried out as described in appendix~D
in {\cite{Bode:1999sm}}. The $\varepsilon$ missing in the table can be
computed from those given. Digits are quoted such that there is at most an
uncertainty of one in the last digit. For $\kappa = 0$ odd corrections vanish.

\end{appendix}

\bibliography{dimerref}           
\bibliographystyle{h-elsevier}

\end{document}